# The role of coherent airflow structures on the incipient aeolian entrainment of coarse particles

Xiao-Hu Zhao[1], Manousos Valyrakis[2,3,*], Thomas Pähtz[4,5], Zhen-Shan Li[6,7]

[1] Division of Environmental Management and Policy, School of Environment, Tsinghua University, Beijing, China.

[2] Department of Civil Engineering, Aristotle University of Thessaloniki, Thessaloniki, Greece.

[3] School of Engineering, University of Glasgow, Glasgow, UK.

[4] Institute of Port, Coastal and Offshore Engineering, Zhejiang University, Zhejiang, China.

[5] Donghai Laboratory, 316021, Zhoushan, China.

[6] College of Environmental Science and Engineering, Peking University, Beijing, China.

[7] The Key Laboratory of Water and Sediment Sciences, Ministry of Education, Peking University, Beijing, China

*Corresponding author: Manousos Valyrakis mvalyra@civil.auth.gr

**Highlights**

-Coherent airflow sweep structures exerting sufficient drag force can set coarse particles in motion.

-A work-based criterion has been established to define the full aerodynamic entrainment of coarse particles.

-The proposed criterion agrees with wind tunnel experimental observations.

**Abstract**

The role of coherent airflow structures capable of setting gravel-size particles in motion is studied theoretically and experimentally. Specifically, a micromechanical model based on energy conservation is proposed to describe the incipient motion of large-particles ranging from rocking (incomplete entrainment) to incipient rolling (full entrainment). Wind tunnel experiments were conducted on an aerodynamically rough bed surface under near-threshold airflow conditions. Synchronous signals of airflow velocities upwind of the test particles and particle displacement are measured using a hot film anemometer and a laser distance sensor, respectively, from which coherent airflow structures (extracted via quadrant analysis) and particle movements are interlinked. It is suggested that the incipient motion of gravel-size particles (rocking and rolling) may result from sufficiently energetic sweep events corresponding to aerodynamic drag forces in excess of the local micro-topography resistance. However, full entrainment in rolling mode should satisfy the presented work-based criterion. Furthermore, using an appropriate probabilistic frame, the proposed criterion may be suitable for describing processes of energy transfer from the wind to the granular soil surface, ranging from the creep transport of gravels to the "mechanical sieving" of mega-ripples, as well as the transport of light anthropogenic debris (such as plastics).

**Plain Language Summary**

The entrainment of particles from a particle bed is a key problem in wind-driven sediment transport. Most existing models define critical entrainment conditions as those at which a bed particle begins to move. Such conditions are typically met during the passing of turbulent flow structures. However, the energy transferred from the structure to the particle may not be sufficient to leapfrog over neighboring bed grains. In this case, the particle will eventually fall back into its bed pocket. Here, we experimentally evaluate an alternative work-based criterion for particle entrainment during the passing of turbulent flow structures. Our results indicate that sweeps of sufficient turbulent energy are predominantly responsible for particle entrainment.

## 1. Introduction

The creep of individual gravel particles is sporadic and intermittent, as demonstrated by tracking the motion of gravel on the crest of gravel mega-ripples in Wright Valley, Antarctica (Gillies et al., 2012) and the Argentinean Puna (de Silva et al., 2013). Such movement is a key process in the formation and evolution of coarse-grained bed forms (Zimbelman et al., 2009; Milana, 2009; Yizhaq et al., 2012). Coarse gravel particles form a surface layer, effectively protecting mega-ripples from wind erosion (Sharp, 1963). However, the sizes of such mega-ripples are limited due to their migration, which may be initiated either by strong wind gusts or by impinging particles transported in saltation mode (Isenberg et al., 2011, Katra et al., 2014, Tholen et al., 2022). In addition to the aforementioned low mobility transport mode of gravel, other potential mechanisms of transfer of wind energy towards performing geomorphic work have been recently identified (de Silva et al., 2013). Particularly in the case where the local micro-topography prevents particles from moving downwind, the resulting rocking motion induced by turbulent airflow might be instrumental for the growth of gravel mega-ripples through "mechanical sieving," a process suggested by de Silva et al. (2013). Thus, a better understanding of the formation and evolution of coarse-grained bed forms, for instance, for modeling purposes, requires detailed investigations of intermittent creep and rocking of gravel particles associated with the unsteady wind.

Traditionally, the unsteady wind is treated as a fully random motion in modeling sediment transport (Anderson, 1987; Spies and McEwan, 2000; Lu et al., 2005; Li et al., 2008, Kok and Renno, 2009). However, studies based on flow visualization have noted that wall-bound turbulence consists of random fluctuations and intermittent and quasi-periodic coherent flow structures (e.g., Kline et al., 1967; Grass, 1971; Adrian, 2007). Such turbulent flow structures (Robinson, 1991) or coherent flow structures (Bauer et al., 2013) hold considerable promise for providing new insights into the mechanics of sediment transport and evolution of bedforms because they generate locally high Reynolds stress, which may account for the heterogeneity of sediment transport across a hierarchy of temporal and spatial scales (Jackson, 1976; Mazumder, 2000; Livingstone et al., 2007).

There is now an increasing understanding that coherent airflow structures may be a major contributor to the aeolian transport of relatively small particles, e.g., sand grains (Bauer et al., 1998; Schönfeldt and von Löwis, 2003; Baas and Sherman, 2005; Chapman et al., 2012, Pähtz et al., 2018). Specifically, flow structures associated with the bursting-like phenomenon (sweeps and ejections) have been of considerable significance for transport by saltation (Sterk et al., 1998; Leenders et al., 2005; Wiggs and Weaver, 2012). However, very few studies report the role of flow structures in the intermittent creep of large-size sediments. This episodic transport of bed material may be inherited from the intermittent character of coherent structures of wind gusts found in natural aeolian environments (e.g., van Boxel et al., 2004; Walker, 2005), which locally disrupt the laminar boundary layer and "push" individual particles in the wind direction.

Generally, because the carrying capacity for water is greater than that for air, the transport of coarse bed material, typically larger than fine gravel, has received more attention in water environments (e.g., Drake et al., 1988; Thorne et al., 1989; Hardy et al., 2009). Investigations show that the incipient rolling of particles is associated with intense sweep events characterized by an optimum combination of fluid force and duration (Valyrakis et al., 2011a; Dwivedi et al., 2011). These sweep events may be associated with the motion of large-scale retrograde vortices (Wu and Shih, 2012) or octant structures (Keylock et al., 2014). Even though significant progress has been made in identifying the dynamic processes leading to the entrainment of large sediment particles (Pähtz et al., 2020), this has not been met in the case of aeolian transport.

This study aims to thoroughly investigate the role of coherent airflow structures on the incipient motion of an individual large (typically larger than fine gravel) particle exposed on a bed surface roughened by same-sized particles. Specifically, an energy criterion is developed to distinguish between the flow structures responsible for rocking (incomplete entrainment) and incipient rolling (full entrainment). This is achieved by considering an energy transfer framework from airflow structures to surface particles for the particle-scale micro-mechanical model. In addition, wind tunnel experiments involving synchronous measurements of local airflow velocity and particle incipient displacements are conducted near the threshold flow condition to verify the proposed model. Considering the potential significance of coherent flow structures on aeolian transport and the lack of comprehensive studies for the aeolian entrainment of gravel particles, this study may be seen as a novel paradigm to comprehensively describe the mechanisms of intermittent geomorphic work of the soil surface due to energetic wind bursts.

## 2. Modeling the role of coherent flow structures on incipient motion

Near-threshold (with regard to particle transport) flow conditions are considered, during which particle motion occurs intermittently. In the case of the simplified micro-mechanical model shown in Fig. 1, it is possible that relatively short-lived but sufficiently energetic coherent flow structures, which energetic retrograde vortices may represent following Wu and Shih (2012), can overcome the intrinsic resistance for this arrangement, setting the exposed particle into motion. At near-threshold conditions, entrainment is a dynamic and sporadically occurring process that cannot be modeled suitably by the average wind strength, with spatially and temporally averaged measures such as frictional velocity (e.g., Greeley and Iversen, 1985; Shao and Lu, 2000). An approximate mechanical model should consider interactions between the flow structures and the individual solid particle.

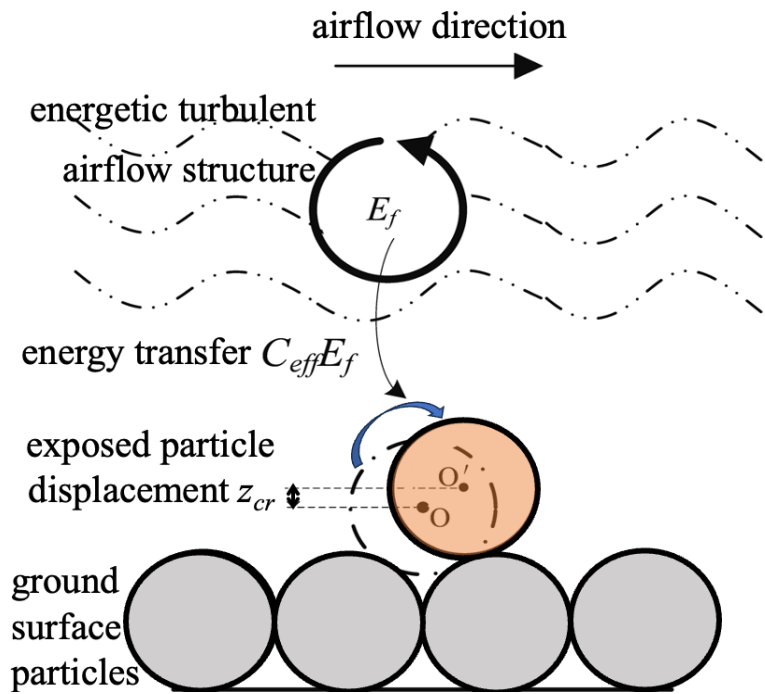

**Fig. 1.** Sketch of a simplified micromechanical model of an exposed particle resting on a rough surface comprising of same-sized particles, showing the energy transferred from the wind burst to the particle,

$C_{eff}E_f$, and the specific elevation ($z_{cr}$) beyond which entrainment is possible by the action of the mean flow forcing alone. $E_f$ is the energy content of the flow structure, and $C_{eff}$ is the energy transfer coefficient.

To investigate the above postulation, the following framework may be considered based on the micro-mechanical arrangement shown in Fig. 1. It is hypothesized that the particle movement from the resting position to a specific elevated position beyond which entrainment is possible by the action of the mean flow forcing alone (O to O′, as shown in Fig. 1) may be initiated by a sufficiently energetic passing flow structure. Based on the conservation of energy, the energy the particle obtains $C_{eff}E_f$ originates from the energy content of the flow structure $E_f$ with a transfer coefficient $C_{eff}$ similar to Valyrakis et al. (2013). This process is achieved by mechanical work of aerodynamic force induced by the flow structure. Then, an appropriate criterion for specifying the threshold level for particle entrainment should link the amount of energy offered by the flow structure with the minimum amount of mechanical work required for particle entrainment, $W_{p,cr}$:

$$C_{eff}E_f > W_{p,cr} \qquad (1)$$

To overcome the energy barrier from O to O′, the mechanical work exerted on the particle should be at least in excess of the increase of gravitational potential ($F_g z_{cr}$; $z_{cr}$ is the vertical displacement from O to O'):

$$W_{p,cr} = F_g z_{cr} \qquad (2)$$

where $F_g [=\pi d^3(\rho_p - \rho_a)g/6]$ is the effective gravity of the particle, $d$ is the particle diameter, $\rho_p$ and $\rho_a$ are, respectively, the density of the particle and air, and $g$ is the gravitational acceleration constant. Rolling friction forces can be two orders of magnitude smaller than the particle's weight (Pöschel et al. 1999). For Eq.(2), any energy losses due to rolling frictional forces are significantly smaller than the uncertainties in calculating the particle and flow event energies and thus can safely be neglected.

Substituting Eq. (2) into Eq. (1) and linking $z_{cr}$ to the particle diameter (*d*) using a parameter *α* ($z_{cr}=αd$) yields:

$$E_f C_{eff} \geq \pi d^4(\rho_p - \rho_a)g\alpha/6 \tag{3}$$

The energy density of individual coherent flow structures may be sufficiently represented with the cube of the local and instantaneous windwise velocity component upwind of particles ($u^3$), neglecting the contribution of any other components (similar to Valyrakis et al. 2013 for water flows). Then $E_f$ can be expressed as the integral of energy density over the time scale (or duration) of the flow structure $T_f$ (Burton et al., 2011):

$$E_f = 0.5\rho_a S \int_{t_0}^{t_0+T_f} u^3 dt \tag{4}$$

where $t_0$ is the occurrence time of flow structures, and *S* ($=0.25\pi d^2$) is assumed to be the spherical particle's cross-sectional area. Substituting Eq. (4) into Eq. (3) yields:

$$C'_{eff} \int_{t_0}^{t_0+T_f} u^3 dt > \omega^2 d\alpha \tag{5}$$

where $C'_{eff}$ ($=C_{eff}/C_d$) is the normalized energy transfer coefficient, $C_d$ is the drag coefficient, and $\omega = \sqrt{4gd(\rho_p - \rho_a)/(3C_d\rho_a)}$) is the fall velocity of particles in air (Durán et al., 2011). Eq. (5) defines a work-based criterion for full entrainment of coarse particles by rolling mode. It links the threshold energy level relevant to flow structures to the properties of solid particles and flow reflected by particle fall velocity *ω* and a specific elevation parameter *α* which may be a random variable at least associated with local pocket arrangements.

The amount of energy offered from flow events to particles, parameterized $C'_{eff} \int_{t_0}^{t_0+T_f} u^3 dt$ according to Eq. (5), is equal to the mechanical work performed on particles neglecting the frictional energy loss. Without considering lift force on particles, the mechanical work is integral to the rate of work of the drag force. The drag force is represented using the classical quadratic velocity parameterization ($u^2$), neglecting the particle's entrainment velocity. Then, similar to Valyrakis et al. (2013), the rate of work (product of drag force and particle windwise velocity $v$) can be parameterized by $u^2v$, and the mechanical work done from the flow structures on the rolling particle can be parameterized using $\int_{t_0}^{t_0+T_f} u^2 v\, dt$ the above analyses to develop $C'_{eff} \int_{t_0}^{t_0+T_f} u^3 dt = \int_{t_0}^{t_0+T_f} u^2 v\, dt$ the equation for estimating the normalized energy transfer coefficient, $C'_{eff}$:

$$C'_{eff} = \int_{t_0}^{t_0+T_f} u^2 v\, dt \Big/ \int_{t_0}^{t_0+T_f} u^3 dt \tag{6}$$

Characteristic values or stochastic distributions of particle sizes and bed configurations can be determined for a specific granular surface. In this case, for a flat, fixed, and well-packed bed surface comprising of unisize spherical particles, parameter $α$ from Eq. (5) is expected to remain invariant and can be defined experimentally. Further, the normalized energy transfer coefficient $C'_{eff}$ defined by Eq. (6) needs to be quantified experimentally. These are investigated in the following.

3. Experiments

3.1. Experimental setup

The experiments were undertaken in an environmental wind tunnel in the State Key Joint Laboratory of Environmental Simulation and Pollution Control, Peking University. The available rectangular section of the wind tunnel is 30 m long, 3 m wide, and 2 m high. The oncoming wind speed ($U$) measured in the middle of the tunnel cross-section, 0.4 m away from the tunnel entrance and 1.2 m above the tunnel floor, can be varied continuously from 0.5 to 20 m/s.

The complexities associated with the variability of natural grains were eliminated using hollow, spherical particles of $d$=40 mm weighing 2.7 g (e.g., a typical table tennis ball made of celluloid) for the test particle. The wind tunnel's bed surface at the test section comprises one layer (bed depth of 4 cm) of uniformly arranged (in a rectilinear fashion), unisize particles of the same properties as the test particle, towards forming a flat and aerodynamically roughened experimental bed surface. The test surface is non-erodible for the tested wind speeds. The fixed bed surface allows for standard estimations of the resistance to entrainment with average shear stress criteria, which, however, have been demonstrated to overestimate the mobility of the bed surface material (Yager et al., 2018; Pähtz et al., 2020). The 4 m long, 2.4 m wide experimental bed test section was located 22.6 m downwind from the tunnel's entrance, where fully developed turbulent conditions were established during the experiments.

Several factors were considered in selecting the particle size. Specifically, the particle size should be a few cm at least to achieve the smallest relative error when considering the resolution of the optical method (laser distance meter with ±0.1 mm accuracy) used in the experiments. Likewise, as a rule of thumb, the chosen particle size should be a fraction (e.g., less than 1/20th) of the effective width of the wind tunnel (which is the portion of the width of the 3 m wide tunnel along which the airflow is fully developed and not influenced by the tunnel walls).

Different particle property combinations were tested. For example, greater size spherical particles might also be able to be entrained even if they might be heavier because they impinge higher into the boundary layer. However, a size and weight combination that would allow the particle to be entrained by the fluctuating coherent flow structures impinging into or generated near the bed surface, rather than because of the average aerodynamic forcing experienced further from the bed surface, is strongly preferred. This would allow the reproduction of the highly intermittent character of aerodynamic entrainment, also seen in the field. Last, the combination of the target particle's weight and size should be such that it would allow for aerodynamic entrainment to be observed for a range of wind tunnel speeds that the turbine can generate. This consideration automatically excludes the ranges of heavier particles if intermittent incipient particle entrainment is to be observed.

If natural gravel with a particle size similar to the test particles were employed in these experiments, a high-speed wind tunnel with wind strength of up to approximately 100 m/s would be required (Batt and Peabody, 1999). Therefore, low-density particles (about 67 times air density) were utilized to allow the running of the experiments in the available low-speed wind tunnel. A similar approach has been used in hydraulic transport for a long period, e.g., Kaftori et al. (1995) used polystyrene particles to investigate the role of coherent wall structures on sediment transport in a laboratory flume. Even though the experimental particles are lighter than natural sediments, they are much heavier than air. Then, the incipient motion of the model is still governed by a self-weight and aerodynamic force similar to natural gravel. Thus, this experimental setup may be appropriate to investigate the elementary dynamics of entrainment for ideally spherical gravel particles on uniform bed surfaces.

Further, it is important to consider the modern challenge of anthropogenic transport of plastics in the environment due to wind. The current analysis focuses on wind tunnel experiments that demonstrate for a first time the assessment of the aerodynamic threshold which accounts for turbulence for the incipient motion of plastic debris, albeit with an idealised spherical shape. These are considering the new dynamic criterion of energy, which fully accounts for the effects of turbulent fluctuations, that scale with the particle dimensions.

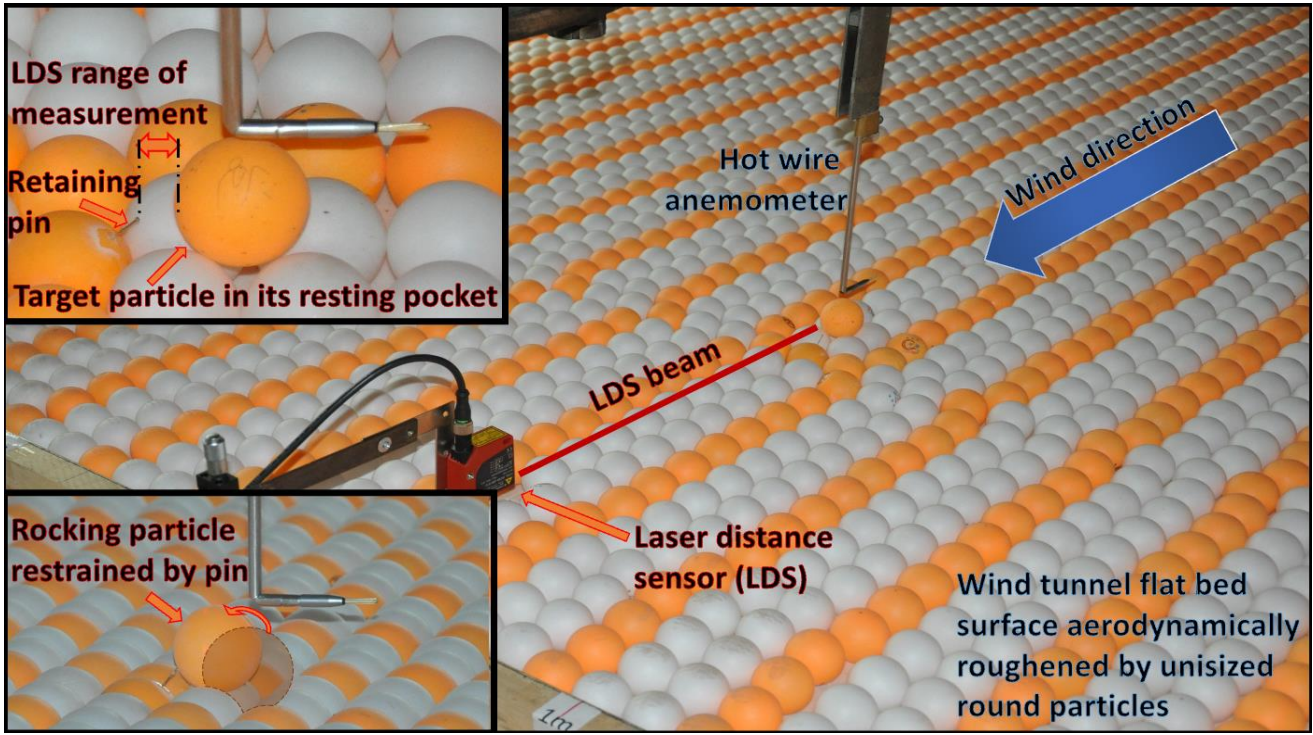

**Fig. 2.** Illustration of the experimental setup used for conducting synchronous particle displacement and airflow velocity measurements. The top left insert, shows the target particle in its resting tetrahedral arrangement, along with an indication of the range of the measurements taken by the laser distance sensor. The bottom left insert, shows the target particle rocking outside of its resting pocket, while its full entrainment downstream is restrained by the retaining pin.

Movements of the particles were recorded by a laser distance sensor (LDS) with 0.1 mm resolution. Calibration revealed that the electrical signal output from the LDS was linear with the distance from the target object to the receptor (similar to Diplas et al., 2010 and Xiao-Hu et al., 2021). The LDS laser beam was fixed to shine at the center of the downstream face of the target particle (Fig. 2). The sensor's measuring range of interest was set from the rest position of the target particle to the top of the supporting downwind particles, where the mobile particle would be considered to have been completely entrained. The LDS measurement range of interest refers to the maximum distance LDS records when the target particle is entrained from the resting position to the position, signifying that full entrainment has been achieved. Continuous measurement records are allowed by using a pin that forces the particle back into its pocket after the fluctuating aerodynamic forcing from advected airflow structures is reduced below the particle's resisting force (which is a component of the particle's weight). For the instances when the aerodynamic forces are too strong, the particle may dislodge further downwind past the pin (Fig. 2). Since only the particle's incipient motion is studied, only records within the LDS measurement range of interest are kept for further analysis. Utilizing such a setup and considering the fixed pocket geometry allows the measured displacement to be decomposed into its windwise and vertical displacement components.

The instantaneous windwise and vertical components of wind velocities ($u$ and $w$, respectively) were measured using a two-dimensional hot-film anemometer positioned at one diameter ($1d$) upwind of the front of the particle and $0.2d$ above the top of the target particle in its resting pocket position (Fig. 2). The velocity measurement location was chosen in order to avoid being too close to the particle, potentially interfering with the local flow field around the particle, or too far, where the measurements would not be representative of the local aerodynamic forcing field around the test particle. Signals of the LDS and the anemometer were sampled synchronously using a multichannel data acquisition card (DAQ) with a frequency of 1000 Hz. Further information on the acquisition and processing of the experimental data is described in the companion supporting information (SI) documentation.

### 3.2. Time-averaged flow conditions in the experiment

The approach velocity ($U$) upwind of the test section in the experiment remained relatively invariant at 8 m/s near the threshold airflow condition. The average windwise wind velocity upwind of the target particle ($\bar{u}$) was approximately 4.8 m/s, and the corresponding particle Reynolds number ($Re_p = \bar{u}d/v$, where $v$ is the kinematic viscosity of air and $v=1.5\times10^{-5}$ m$^2$/s) was approximately 13,000. At such a high Reynolds number, the airflow around particles is fully turbulent, and the drag force results from the asymmetry of pressure between the two sides of the particle. This regime is consistent with that of natural gravels, considering their large size, density, and the resulting high fluid threshold.

The conditions of the simulated atmospheric boundary layer are shown in Fig. 3, where the profile of the average windwise wind velocity at the measuring location followed the typical logarithmic law 10 mm above the tops of the particles comprising the bed surface (Fig. 3a). The turbulence intensity is decreased with the height from the bed surface, with a value of up to 20% near the bed surface (Fig. 3b). The inertial subrange at frequencies >10 Hz for windwise velocity components upwind of the particle is characterized by a typical Kolmogorov −5/3 power law (Fig. 3c). The energy-containing range at frequencies <10 Hz, scaled with a −1 spectral slope, is ascribed to large-scale eddies in which the energy production and cascade energy transfer coexist. These results are in accordance with a naturally occurring atmospheric boundary layer.

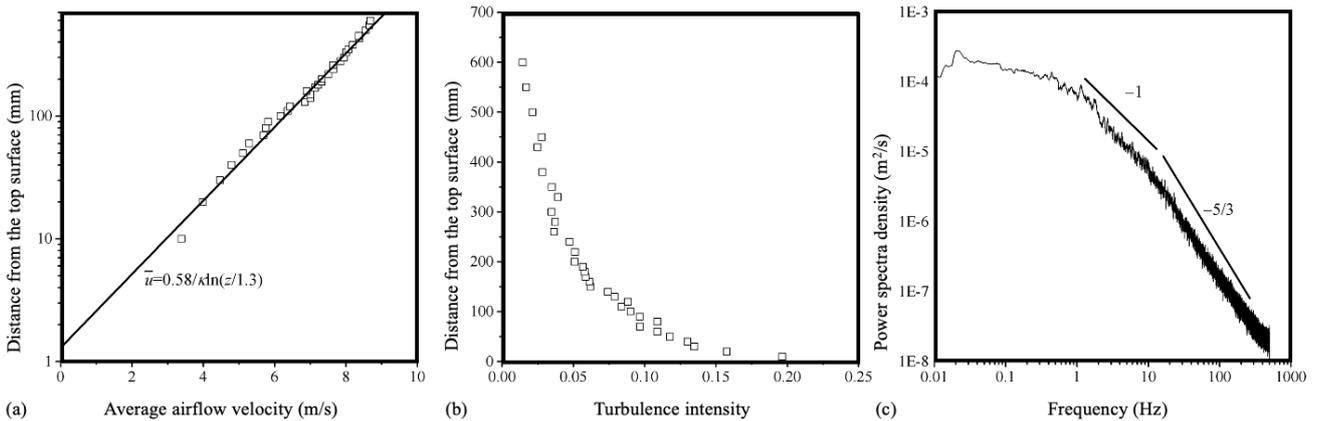

**Fig. 3.** (a) Windwise wind velocity profile, (b) turbulence intensity profile, and (c) $u$-power spectra at the measuring location at $U=8$ m/s. The zero-reference level is set to the tops of the rough bed's spherical particles. $\kappa$ (=0.4) is the Karman constant.

## 3.3. Data processing

Coherent flow structures associated with near-surface bursting processes were extracted using the quadrant technique given by Lu and Willmarth (1973). Briefly, the coherency of local airflow is identified by the magnitude of kinematic Reynolds stress (negative covariance of $u'$ and $w'$, $-u'w'$) based on a subjective threshold $|u'w'| \geq H u_{rms} w_{rms}$, where $u_{rms}$ and $w_{rms}$ are the root-mean-square values of $u'$ and $w'$ signals, respectively. The parameter $H$ in this study was set to 1, following the detailed investigation of Bogard and Tiederman (1986). Stress events of high magnitude, which indicate the presence of coherency in the local flow, are decomposed into four categories according to $u'$ and $w'$ quadrants: i) quadrant 1 (Q1) outward interactions ($u'>0$, $w'>0$), ii) quadrant 2 (Q2) ejections ($u'<0$, $w'>0$), iii) quadrant 3 (Q3) inward interactions ($u'<0$, $w'<0$), and iv) quadrant 4 (Q4) sweeps ($u'>0$, $w'<0$). Stress events below the threshold are called "hole events" comprising noncoherent (or random) flow components.

Following the above procedure, synchronous time series of quadrant and displacement events are acquired and subsequently analyzed. Signatures of quadrant events are dominated by Q4 and Q2 events, which occur with a frequency of approximately 2 counts/s as shown in Table 1. Q1 and Q3 events infrequently occur with a frequency of 0.2 and 0.3 counts/s, respectively. Correspondingly, the average value of interval time ($T_i$) of Q1 and Q3 events is much higher than that of Q2 and Q4 events. The location of the anemometer probe may influence the detection of the occurrence of quadrant events. According to Wu and Shih (2012), if the probe is positioned relatively far (e.g., one particle diameter similar to our experimental setup) from the upwind front of the target particle, the quadrant technique is inclined to predict more Q4 events than Q1 events. This might be true considering the body of flow structures is three-dimensional, and a single-point velocity measurement can only grasp the local features of flow structures. The displacement events occur much less frequently than quadrant events (see Table 1). This suggests that most quadrant events are ineffective for particle incipient motions if the latter is supposed to result from quadrant events. The average value of the duration of quadrant events ($T_f$) is smaller than the average value of ascending duration of particles ($t_a$), defined as the duration of downwind movements from the resting location to the peak for rocking motions and to the location where full entrainment is achieved for incipient rolling. This may imply that quadrant events only contribute a portion of downwind displacements of particles.

**Table 1** Time-averaged quadrant and displacement event parameters, including frequency, duration, and interval time (until the stochastically occurring full entrainment by rolling, when the test particle is fully removed from its resting pocket).

|  | Quadrant events | | | | Displacement events | |
|---|---|---|---|---|---|---|
|  | Q1 | Q2 | Q3 | Q4 | Rocking | rolling |
| Frequency (counts/s) | 0.2 | 2 | 0.3 | 2 | 0.1 | 0.006 |
| Mean duration (s) | 0.03 | 0.06 | 0.03 | 0.05 | 0.12 | 0.36 |
| Mean interval time (s) | 4.8 | 0.49 | 3.5 | 0.48 | 10.3 | – |

## 4. Results

### 4.1. Characteristics of quadrant events responsible for particle incipient motion

The test particle was observed to roll intermittently and to rock preceding the incipient rolling under our experimental conditions, similar to the behavior of natural gravels (de Silva et al., 2013). To examine which quadrant event was responsible for each recorded particle motion, comparisons were made of the occurrence of rocking and rolling to quadrant events. It is found that the rocking motions can be caused by Q4 or Q1 events alone (e.g., event A in Fig. 4a and event B in Fig. 4b, respectively) or by sequences of Q1 events followed by Q4 events (e g., event C and event D in Fig. 5a). The occurrence of incipient rolling similarly has an intimate relationship with sweep events as event E shown in Fig. 5b. Descriptive statistical indices for all movements detected in the experiments, such as counts of rocking (incomplete particle motions) and rolling (full entrainments) and the type of flow structure responsive for this particle response, are reported in Table 2. It is worth noting that the majority of complete and partial particle displacements, more than 85%, occur due to Q4 events alone. Thus, sweeps are the most relevant quadrant events for the incipient motion of coarse particles. These results agree with some investigations in water flows (e.g., Hofland and Booij, 2004; Detert et al., 2010). However, they are distinct from what Wu and Shi (2012) reported, suggesting that Q1 events dominate the instant of particle entrainment. This may be because the measuring location of flow velocity in Wu and Shi (2012) was very close to the target particle ($1/8d$), and Q4 events were not readily detected, as we have also mentioned.

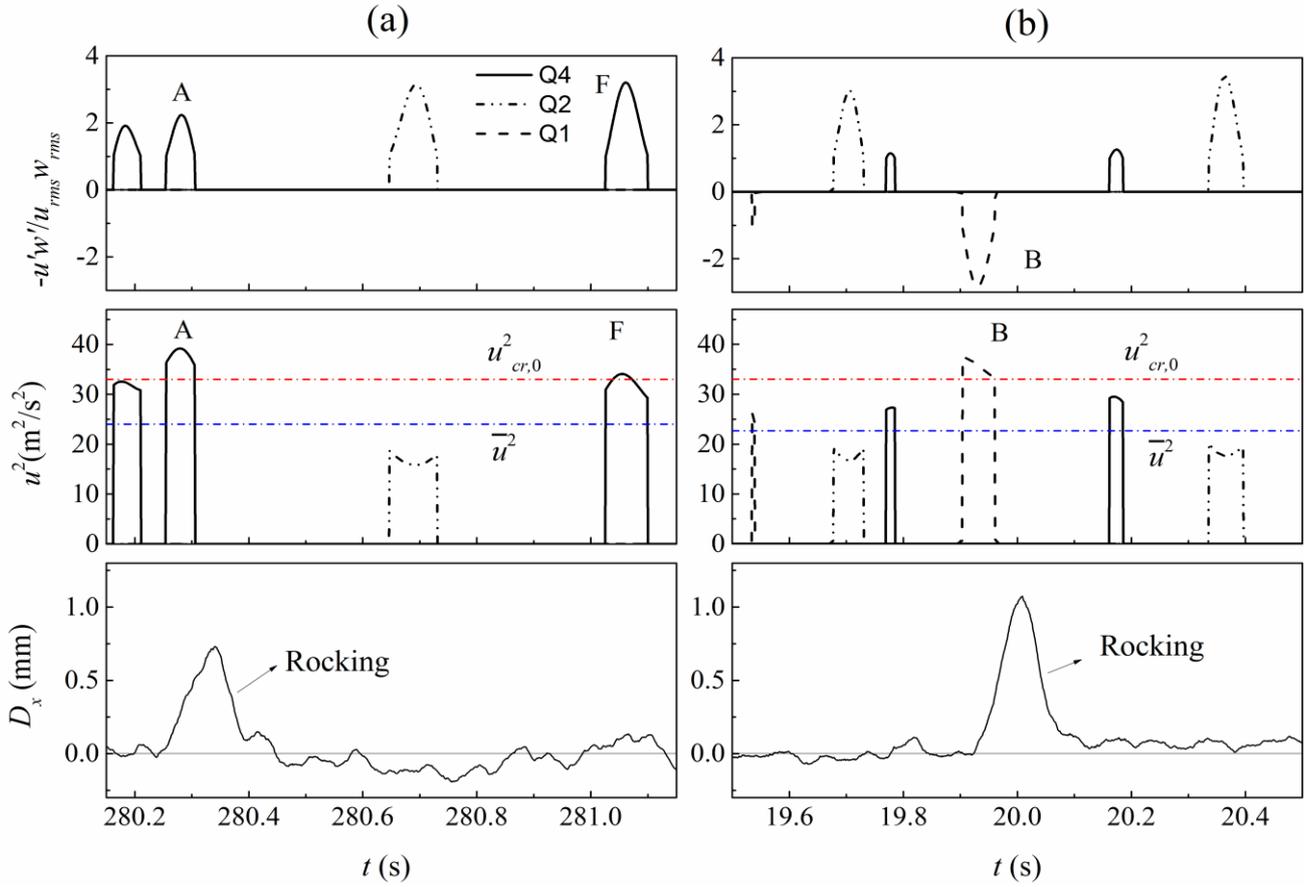

**Fig. 4.** Representation of 1s histories of quadrant events (described respectively by normalized kinetic Reynolds stress, $-u'w'/u_{rms}w_{rms}$, and quadratic windwise airflow velocities, $u^2$) and windwise particle displacement ($D_x$) to demonstrate rocking motions caused by (a) Q4 events (event A) and (b) Q1 events (event B) alone. The subthreshold airflow condition is indicated by the average windwise velocity below the threshold level, $\bar{u} < u_{cr,0}$.

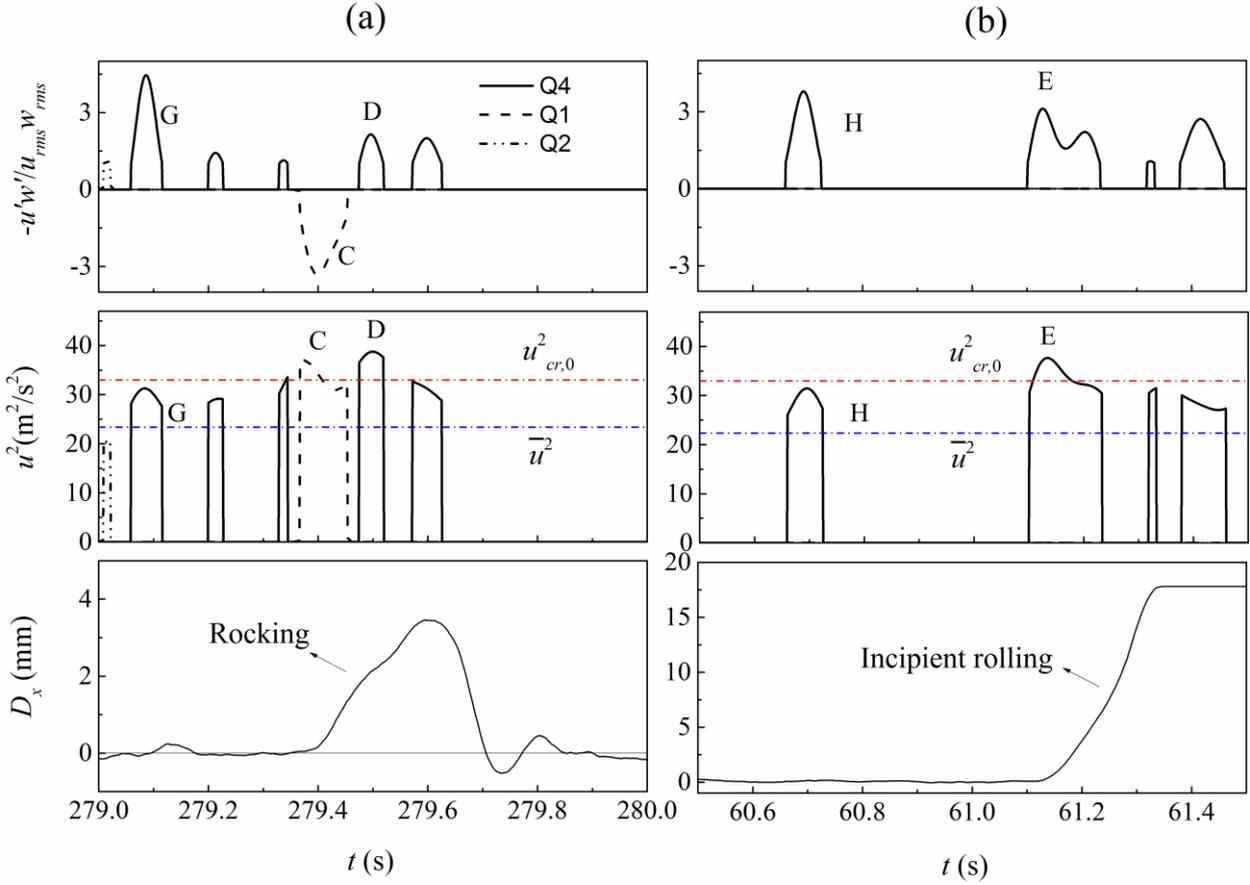

**Fig. 5.** Representation of 1s histories of quadrant events (described respectively by normalized kinetic Reynolds stress, $-u'w'/u_{rms}w_{rms}$, and quadratic windwise airflow velocities, $u^2$) and windwise particle displacement ($D_x$) to demonstrate (a) rocking motions caused by a sequence of Q1 events following by Q4 events (event C and event D, respectively) and (b) the incipient rolling caused by Q4 events. The subthreshold airflow condition is indicated by the average windwise velocity below the threshold level, $\bar{u} < u_{cr,0}$.

**Table 2** Percentage of rocking and rolling resulting from quadrant events

| Movements | Counts | Q1 | Q4 | Q1+Q4* | Q2 | Q3 | Hole events |
|---|---|---|---|---|---|---|---|
| Rocking | 291 | 4.1% | 86.9% | 1.7% | 0 | 0 | 7.2% |
| Incipient rolling | 18 | 0 | 88.9% | 11.1% | 0 | 0 | 0 |

* "+" denotes that particle movements are caused by sequences of Q1 events followed by Q4 events

Sweep events resulting in particle movements are characterized by randomness, and probabilistic models can describe their features. Such as, the interval time between events follows a negatively exponential distribution (Fig. 6a) which means the trend of intensive emergence of energetic sweep events. The duration of events $T_f$ and event energy ($\sim \int_{t_0}^{t_0+T_f} u^3 dt$) follow extreme value theory distribution models (Fig. 6b and Fig. 6c, respectively) characterized by a right, long tail implying that high-energy, long events are scarce.

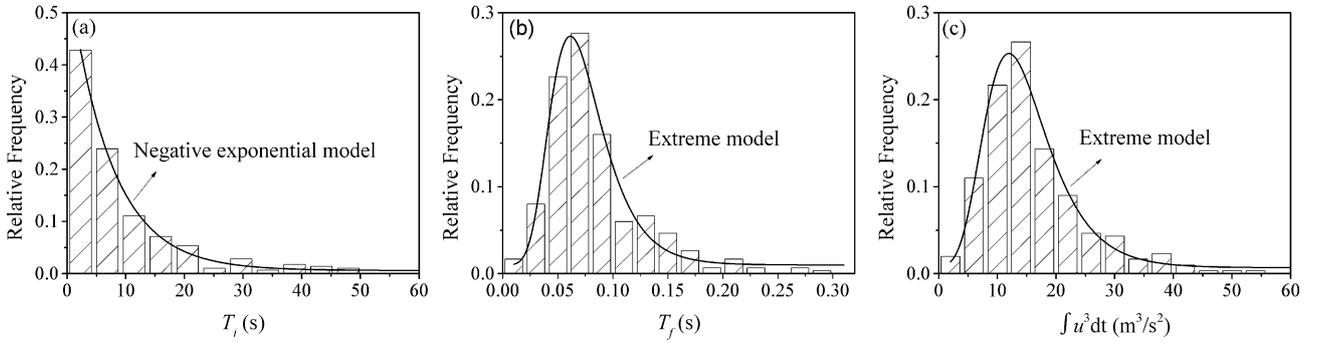

**Fig. 6.** Frequency distributions of (a) interval time $T_i$, (b) duration $T_f$ and (c) energy content ($\sim \int_{t_0}^{t_0+T_f} u^3 dt$) of sweep events resulting in particle movements.

Although sweep events are shown to be the relevant type of flow structures for particle incipient motion, inspections of synchronous time histories of particle displacements and quadrant events similar to Fig. 4 and Fig. 5 reveal that many sweep events do not lead to movements even though they may induce high stress (e.g., event F in Fig. 4a, event G in Fig. 5a and event H in Fig. 5b). It appears that only those events capable of inducing higher drag force than the initial surface resistance level (parameterized by the fluid threshold, $u^2_{cr,0}$) are effective for particle movements. $u^2_{cr,0}$ is derived from Bagnold's force model using the mean drag coefficient $C_d$ (=0.76) from Schmeeckle et al. (2007), where spheres in near-bed turbulent flows were shown to have a larger drag coefficient than that for a sphere settling in still water at similar particle Reynolds numbers.

This speculation is further confirmed by results shown in Fig. 7. It can be seen that the peak drag force (parameterized by the peak of quadratic windwise airflow velocities during the occurrence of flow events, $u^2_{f,p}$) of most of the sweep events relevant to particle movements, more than 80%, is in excess of $u^2_{cr,0}$. The data points below $u^2_{cr,0}$ may attribute to sweep events owning local drag coefficients greater than the average value. Those events are sufficient to induce peak drag force exceeding the local surface resistance level despite their relatively low peak flow velocities. The above analyses support there is a force criterion ($u^2_{f,p} > u^2_{cr,0}$) for sweep events exciting rocking and rolling.

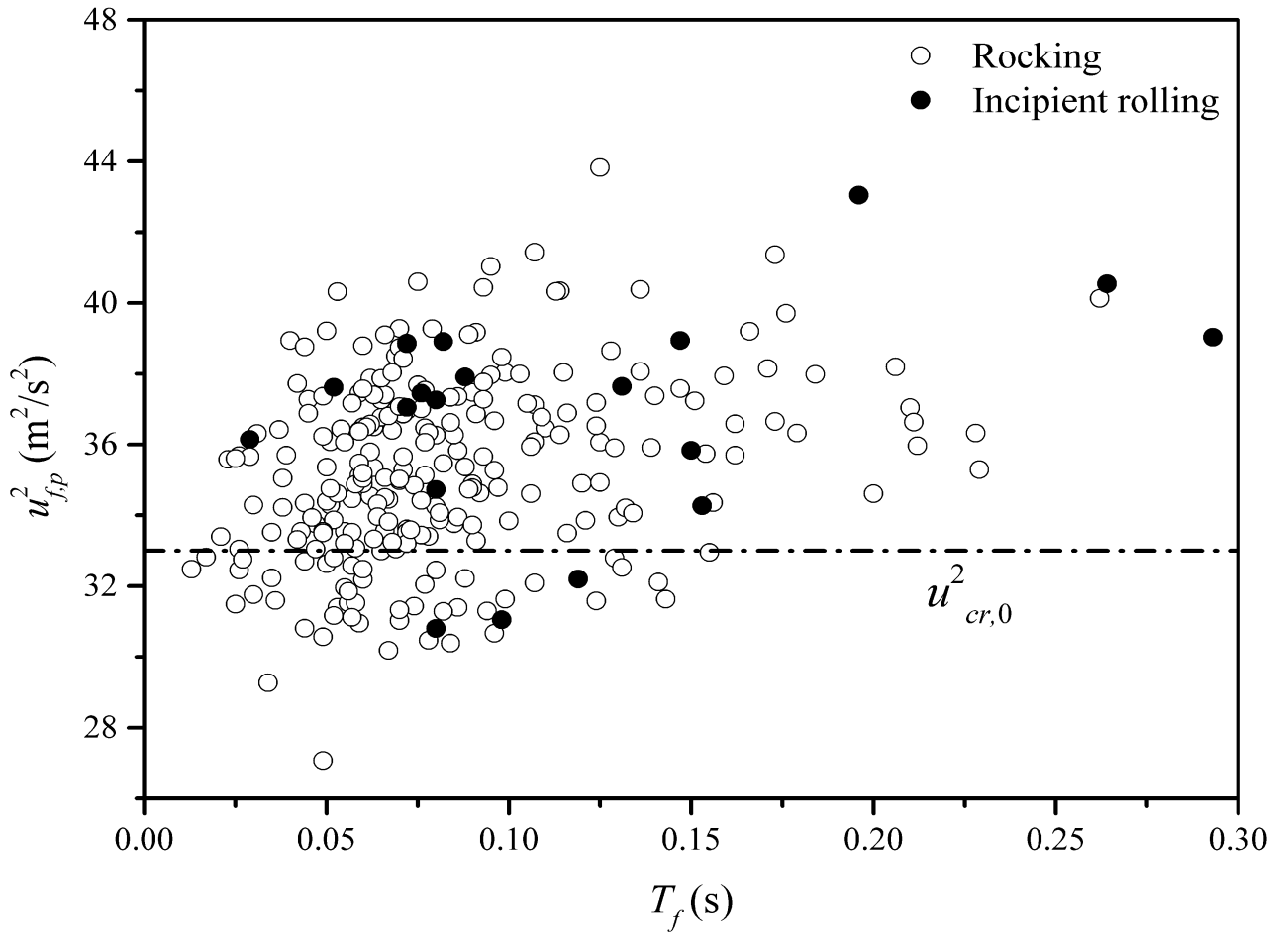

**Fig. 7.** Plot of the peak drag force ($\sim u^2_{f,p}$) against the duration ($T_f$) of sweep events resulting in particle movements to demonstrate the force criterion, $u^2_{f,p} > u^2_{cr,0}$.

However, the force criterion $u^2_{f,p}>u^2_{cr,0}$ is no longer a sufficient condition for sweep events fully dislodging particles. It is shown in Fig. 7 that the magnitude of peak drag force ($\sim u^2_{f,p}$) of sweep events makes little sense in distinguishing incipient rolling from rocking motions. Thus, a force criterion alone is insufficient to define full entrainment, as suggested by previous studies (Diplas et al., 2008; Valyrakis et al., 2010).

### 4.2. Parameters associated with the theoretical energy criterion

The value of the critical vertical displacement $z_{cr}$ ($=\alpha d$) for sweep events is acquired from the vertical displacement measurements caused by sweep events ($z_f$) in the ascending phase of rocking motions, as shown in Fig. 8, which gives $\alpha = 0.053$. Once the particle performs a greater vertical displacement, the mean flow forcing, greater than the reduced resistance at the displaced location of greater exposure to the flow, is enough to entrain the particle completely (similar to Valyrakis et al., 2011b).

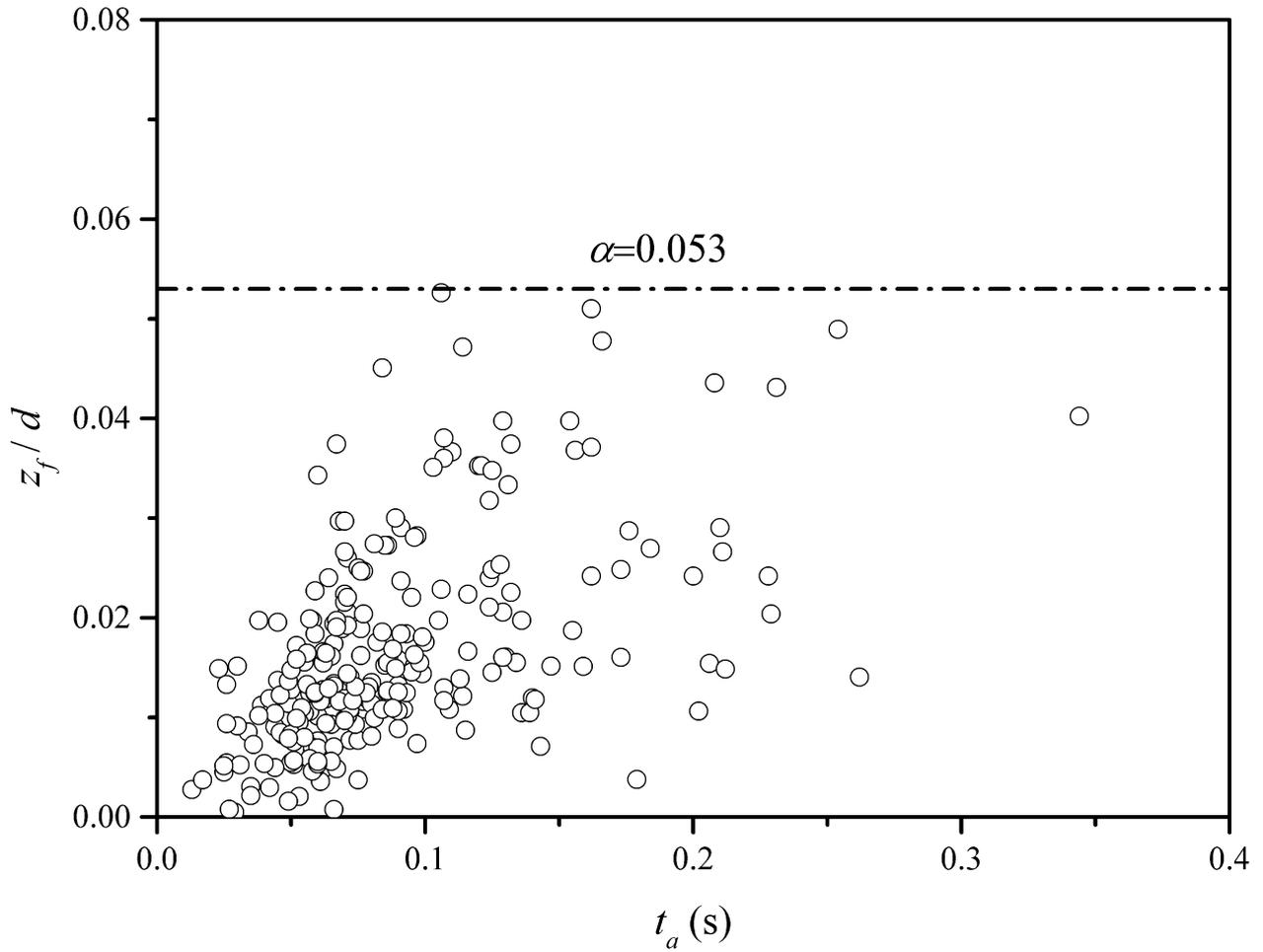

**Fig. 8.** Plot of vertical displacements of particles resulting from sweep events ($z_f$) in the ascending phase of rocking motions against ascending duration ($t_a$) of the particle showing the critical vertical displacement required by sweep events to dislodge particles fully, $z_{cr}$ (=$\alpha d$).

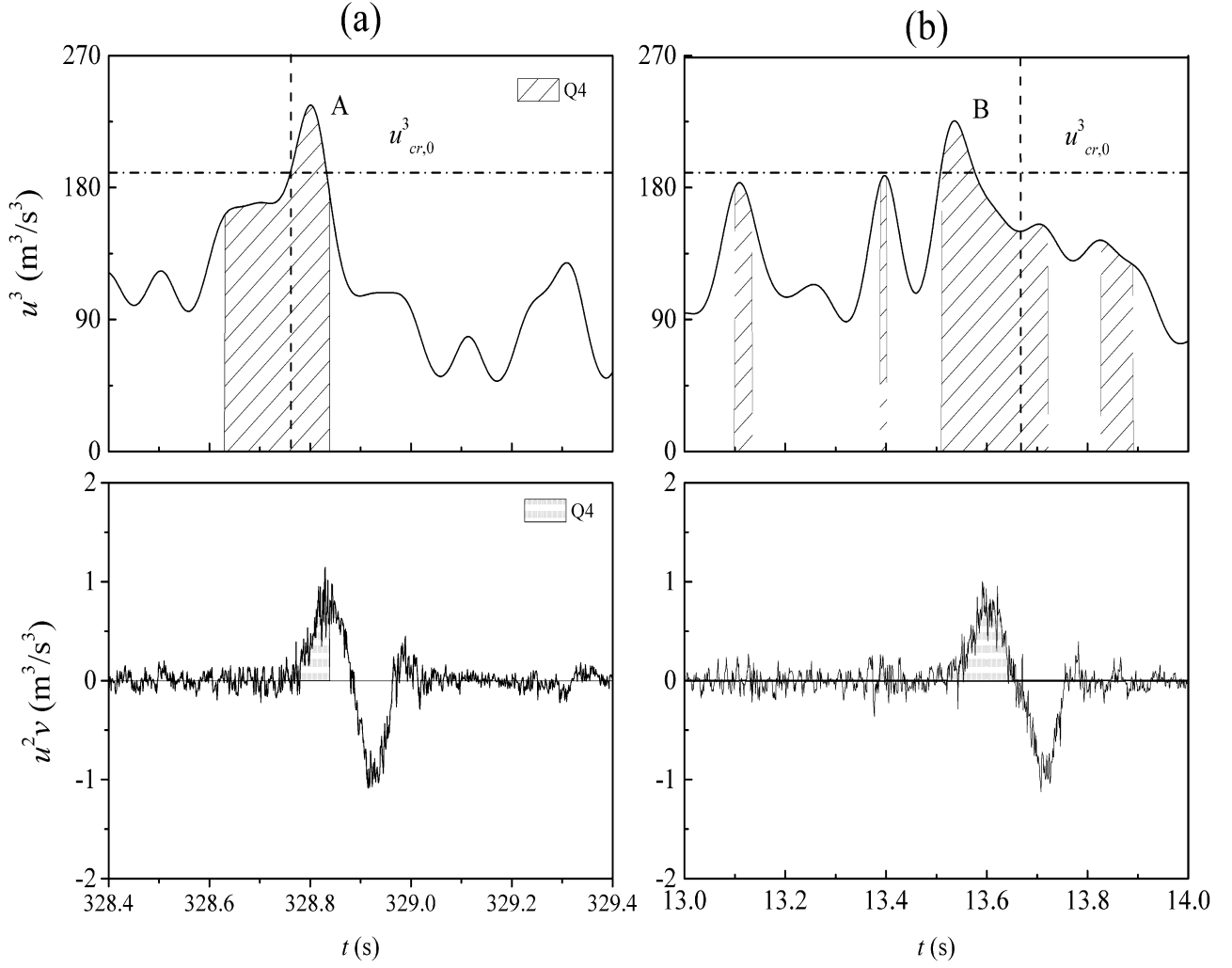

**Fig. 9.** Representation of 1s histories of the rate of work of air drag ($\sim u^2v$) and energy density ($\sim u^3$) near the instance of rocking motions showing the calculation of normalized energy transfer coefficient of sweep events, $C'_{eff}$, which equals the work of air drag (area of grey shadow) divided by corresponding event energy (area of stripe shadow). These examples also demonstrate the low value of $C'_{eff}$ caused by relatively long sweep events where only a fraction of event energy is available to the particle. Such as, the energy content of event A at the left of the vertical dashed line and event B at the right of the vertical dashed line cannot be utilized by particles for downwind ascending motions.

The values of normalized energy transfer coefficient $C'_{eff}$ ($=C_{eff}/C_d$) defined by Eq. (6) are obtained from synchronous time histories of energy density ($\sim u^3$) and rate of work of drag force ($\sim u^2v$). For example, $C'_{eff}$ of sweep events shown in Fig. 9 (event A and event B) equals the work of air drag (the area of grey shadows) divided by corresponding event energy (the area of stripe shadows). It is noted that there may be multiple sweep events transferring energy to the same displacement events. In this case, each event is assigned an individual value of $C'_{eff}$. $C'_{eff}$ is distributed randomly following a lognormal distribution model (Fig. 10a). However, a general trend is observed in which relatively distinct $C'_{eff}$ values exist for rocking and incipient rolling (Fig. 10b). Generally, $C'_{eff}$ is from 0.0008 to 0.005 for rocking and from 0.005 to 0.018 for incipient rolling. This suggests *that* $C'_{eff}$ is an important parameter to distinguish rocking from rolling. For sweep events entraining particles fully, $C'_{eff}$ should be at least greater than 0.005.

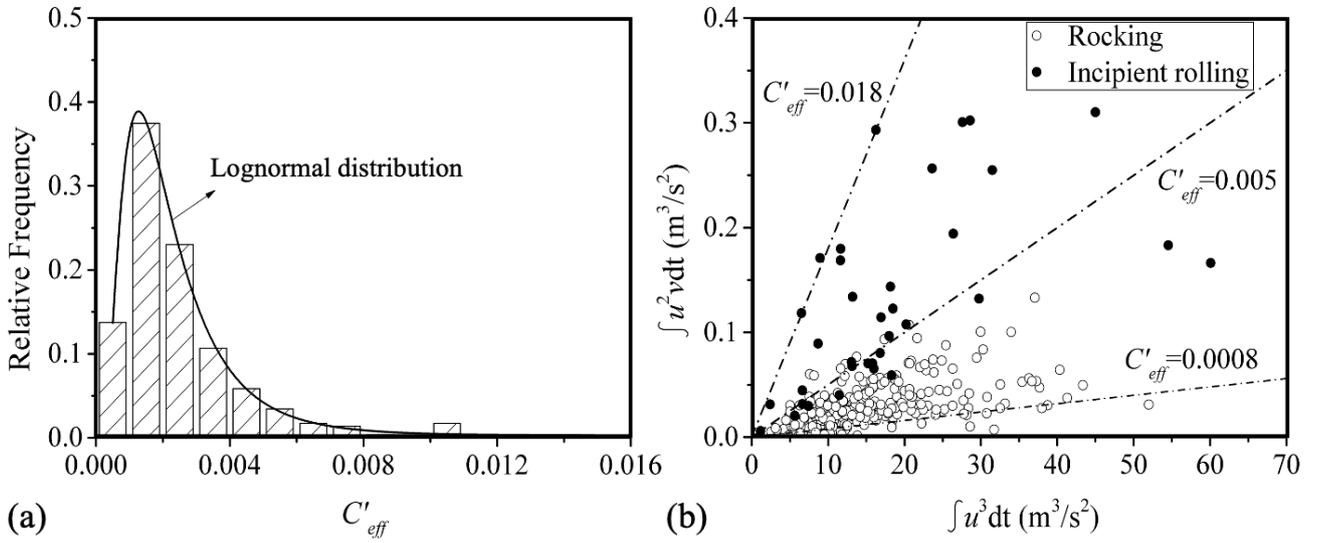

**Fig. 10.** (a) Frequency distribution of normalized energy transfer coefficient, $C'_{eff}$ and (b) plot of event energy ($\sim \int_{t_0}^{t_0+T_f} u^3 dt$) against work of air drag ($\sim \int_{t_0}^{t_0+T_f} u^2 v dt$) of sweep events to demonstrate distinct $C'_{eff}$ for rocking and incipient rolling.

The relationship between the magnitude of $C'_{eff}$ and structures of flow events is elusive. However, if sweep events are long but only a fraction of event energy can be available for particle downwind movements, $C'_{eff}$ is generally low. Two representative instances are provided in Fig. 9 (event A and event B). In the first context (event A), the energy content in the head of events (flow energy on the left of the vertical dashed line) cannot be utilized by particles because these flow components cannot induce sufficient drag force exceeding the initial surface resistance level. In the second context (event B), the energy content in the tail of events (flow energy on the right of the vertical dashed line) is not available because these flow components cannot fight the temporal surface resistance level and then fails to continue to offer energy for pushing the particle up downwind.

### 4.3. Comparison of predictions to measurements

Sweep events transferring energy to particles were extracted for analysis. If several sweep events transfer energy to the same displacement events, the corresponding energy obtained by the particles is the sum of energy offered by each event. Comparisons of threshold energy levels defined by Eq. (5) to experimental measurements are presented in Fig. 11. The theoretical threshold level successfully accounts for 94% of rolling and 99% of rocking motion. This result supports our hypothesis that sweep events have a threshold energy level leading to full entrainment.

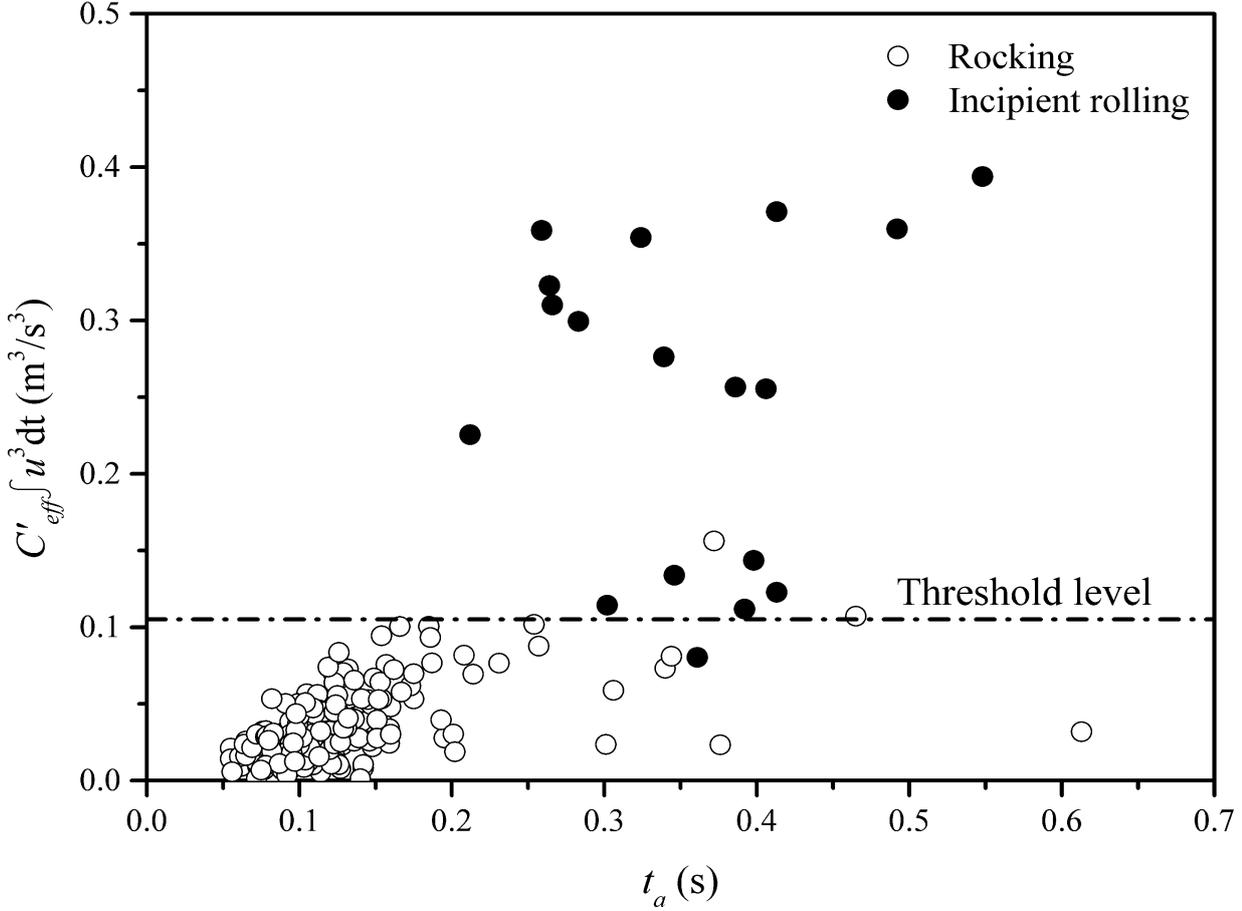

**Fig. 11.** Plot of energy offered by sweep events ($\sim C'_{eff} \int_{t_0}^{t_0+T_f} u^3 dt$) to particles against ascending duration of particles ($t_a$). Parameter values in Eq. (5): $\rho_p$=80.8 kg/m$^3$, $d$=40 mm, $\rho_a$=1.2 kg/m$^3$, $C_d$=0.76, $g$=9.8 m/s$^2$ and $\alpha$=0.053. The values of normalized energy transfer coefficient $C'_{eff}$ (=$C_{eff}/C_d$) defined by Eq. (6) for each event are obtained from synchronous time histories of energy density ($\sim u^3$) and rate of work of drag force ($\sim u^2 v$).

### 4.4. Modes of achievement of full entrainment

The energy offered by sweep events to particles ($E_f C_{eff}$) relates positively to the energy transfer coefficient ($C_{eff}$) and duration of events ($T_f$). Thus, the combination of magnitudes of $C_{eff}$ and $T_f$ can lead to various modes of achievement of full entrainment. For example, a long event with high $C_{eff}$ naturally entrains particles easily, similar to event A as shown in Fig 12. However, relatively short events but owing high $C_{eff}$ or long events with low $C_{eff}$ (e.g., event B and event C in Fig. 12, respectively) be equally competent for full entrainment.

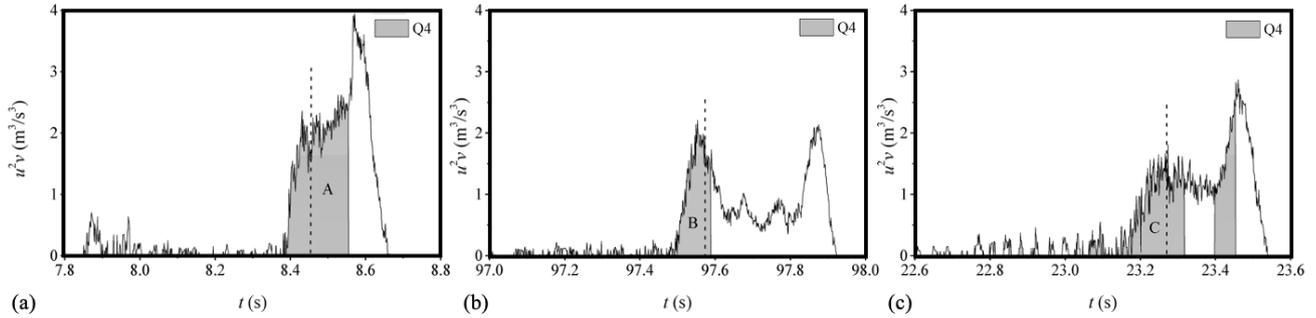

**Fig. 12.** Representation of 1s histories of the rate of work of air drag to demonstrate the single-event mode of full entrainment (incipient rolling): (a) the full entrainment is caused by a relatively long ($T_f$=0.196 s) sweep event (event A) with high energy transfer coefficient ($C'_{eff}$=0.0069); (b) the full entrainment is caused by a relatively short ($T_f$=0.088 s) sweep event (event B) and high energy transfer coefficient ($C'_{eff}$=0.0066); (c) the full entrainment is caused by a long sweep event ($T_f$=0.264 s) sweep event (event C) and relatively low energy transfer coefficient ($C'_{eff}$=0.0028). The vertical dashed line denotes the moment when the threshold energy level is achieved.

Except for the single-event mode mentioned above, a portion of rolling motions (approximately 40%) is achieved by the combined actions of two successive sweep events, as shown in Fig. 13. In this mode, the first sweep event suffices for setting a particle into motion; however, alone it cannot supply sufficient energy to entrain the particle fully. Instead, it may push the particle to a an elevated location of reduced resistance (e.g. see event A, in Fig. 13 a). Subsequently, the shortage of energy is compensated by a trailing aerodynamic event (e.g. see event B, in Fig. 13 a). In some situations, the following aerodynamic event, similar to event D in Fig. 13b, might be able to offer sufficient energy for full entrainment on its own (e.g. not requiring the event C preciding).

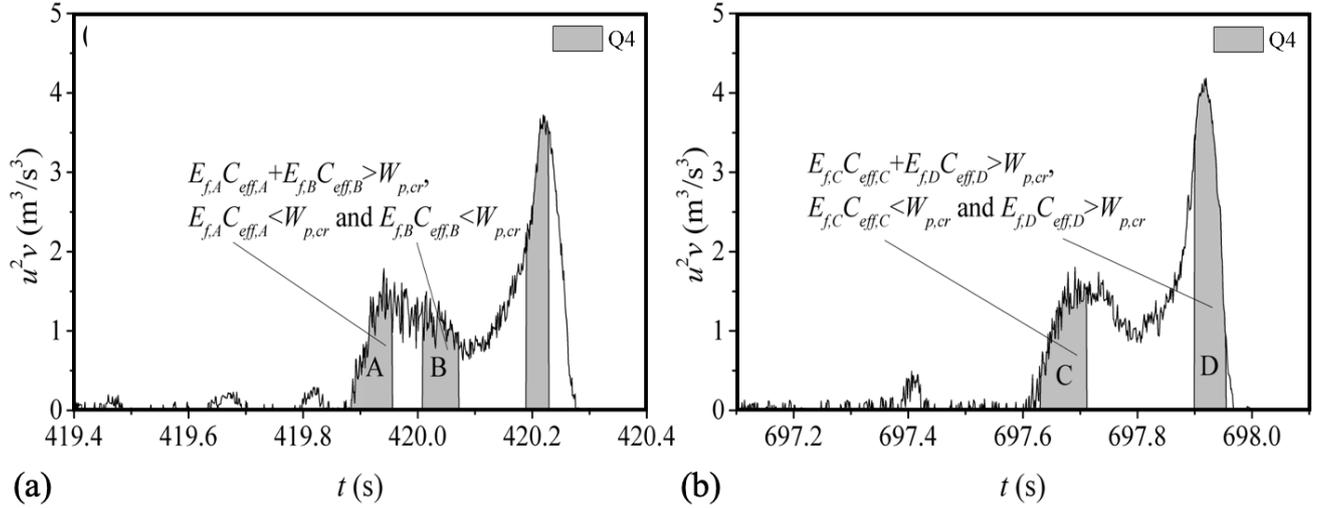

**Fig. 13.** Representation of 1s histories of the rate of work of air drag to demonstrate the multi-event mode for full entrainment (incipient rolling): (a) the full entrainment is caused by two sweep events (event A and event B), and any of them alone cannot supply sufficient energy to particles; (b) the full entrainment is caused by two sweep event (event C and event D), and the event D alone can supply sufficient energy to particles.

Considering how to cluster the impact of such aerodynamic events together, towards a single full particle entrainment, can be seen as a methodological refinement that can be considered for future analysis, along with considering other forcing contributions than direct aerodynamic drag. It is clear that those events should have relative temporal proximity, so that their effect does not fade out with the particle falling back into its initial resting pocket. At least, the time from the end of the previous event to the onset of the following event should be shorter than the duration of ascending motions of particles. For example, measurements indicate that the ratio of the time between two sweep events transferring energy to the same displacement events to the corresponding ascending duration is from 0.11 to 0.41 for incipient rolling and from 0.08 to 0.52 for rocking. Another hint refers to the occurrence of sweep events in airflow, which may inherently have a grouped pattern similar to water flow (Best, 1992).

## 5. Discussion

Aeolian transport, especially for large-size particles, is intrinsically characterized by intermittency posing great challenges to the explanatory ability of the traditional viewpoint based on the mean flow. On the other hand, the event-based flow viewpoint brings up a promising methodology to solve the issue of intermittency because it isolates strong components from the turbulent flow by employing a certain standard, and these isolated components may be the stockholders of local transport phenomena.

### 5.1. The utility of the quadrant method and the energy criterion

This study focuses on the role of coherent airflow events on the incipient motions of individual gravel particles using the quadrant technique (Lu and Willmarth, 1973), which detected flow events based on signs of fluctuation velocity and a threshold of Reynolds stress. The results suggest that the incipient motions of gravel-size particles involving rocking and incipient rolling are attributed to sweeps and outward interactions, both of which have positive windwise fluctuation velocities. Furthermore, on flatbed surfaces, time histories of the windwise flow velocities instead of vertical flow velocities or stress are coupled well with the time histories of the drag force (Schmeeckle et al., 2007). This implies that only sweeps and outward interactions have the potential to induce a higher instantaneous drag force than the initial surface resistance level under subthreshold flow conditions where the time-averaged drag force is lower than the initial surface resistance level.

The quadrant technique provides a rough empirical approach for identifying flow events responsible for particles' incipient entrainment. It classifies high-stress events into four types, with clearly sweeps contributing the majority of cases leading to coarse particles' incipient entrainment, as shown in Table 2. However, quadrant analysis only considers the magnitude of aerodynamic forces thus being unable to establish which amongst the sweeps would be effective to fully dislodge particles downwind, as demonstrated in Fig. 4 and Fig. 5. This study similar to Valyrakis et al. (2013), considers that only quadrant events of sufficient aerodynamic force and energy intensity can lead to complete aerodynamic entrainment. The force criterion ensures that sweep events can set particles into motion. However, alone it cannot determine whether the resulting motion is a rocking event (incomplete entrainment) or an incipient rolling (full entrainment) because the incipient rolling occurs only when a critical vertical displacement is achieved. The presented energy criterion links the critical vertical displacement to the energy content of the quadrant events, required to perform the critical mechanical work. It synthesizes the features of sweep events (e.g., energy transfer coefficient $C_{eff}$ and duration $T_f$), both having influences on incipient rolling (full entrainment) of particles and successfully distinguishing them from rocking motions (incomplete entrainment).

The theoretical frame of particle entrainment has seen innovative developments in recent years (e.g., Diplas et al., 2008; Celik et al., 2010; Valyrakis et al., 2010; Valyrakis et al., 2011b; Valyrakis et al., 2013). These frameworks are based on the transfer of flow energy or momentum and are distinct from the traditionally employed Shields or Bagnold's criteria for identifying the flow conditions for intermittent entrainment of particles in turbulent flows. Such flow events are called energetic flow events (Valyrakis et al., 2013), similar to flow components above the dash-dot line ($u^3_{cr,0}$) in Fig. 9. They are characterized by all components of the flow events capable of inducing sufficient drag force in excess of the surface resistance level. It may be interesting to explore the relationship between energetic flow events and sweep events, considering both can entrain particles. As visualized in Fig. 9, a simple analysis reveals that the energetic flow and sweep events resulting in particle movements are partially overlapping. However, the definition of sweep events only implies a flow stress criterion and lacks any connection to particle properties. Therefore, the force criterion developed using the windwise flow velocity should be presented to help distinguish which sweep events can induce sufficient drag force to exceed the initial surface resistance level. In addition, energetic sweep events can involve ineffective aerodynamic forcing components that do not transfer energy to particles, as event A and event B are shown in Fig. 9. Considering the above, the energetic flow events criterion is superior to the sweep events method in assessing incipient particle entrainment because it has embedded the initial condition ($u^2 > u^2_{cr,0}$) of initiation of particle movements, as well as considering the duration of aerodynamic forcing.

**5.2. Modeling particle rocking and rolling**

The rocking and rolling of gravel particles are relatively new topics for aeolian transport and amongst foundational dynamic processes for forming and evolving aeolian coarse-gravel mega-ripples. Specifically, both movements can be initiated aerodynamically without requiring the impact of fine materials such as sand grains (de Silva et al., 2013). However, previous numerical models for mega-ripples (Yizhaq, 2004; Yizhaq, 2008) generally neglect these aerodynamic processes, consider the entrainment of coarse particles by the impact of sand grains, and naturally cannot provide comprehensive descriptions for mega-ripples. This study is the first attempt to investigate the interaction between coarse particles and turbulent airflow. The presented force and energy criteria could be used to estimate the occurrence probability ($P$) of intermittent rocking and rolling in an appropriate probabilistic framework (e.g., extreme value theory suggested by Valyrakis et al., 2011b), considering both $E_f$ and $C'_{eff}$ are random variables and large value of them conducive to full entrainment are rare events implying by their long-tail probability distribution modes (see Fig. 6c and Fig. 10a). For example, the rocking probability can be expressed as the product of $P(u^2_{f,p}>u^2_{cr})$ and $P(E_f C_{eff}<W_{p,cr})$. Similarly, the rolling probability is the product of $P(u^2_{f,p}>u^2_{cr})$ and $P(E_f C_{eff} >W_{p,cr})$. If the number density of sweep events in a specific region has been obtained, the number density of fully entrained particles in this region, which may be necessary for modeling mega-ripples, is the product of rolling probability and number density of sweep events. However, this idea is only available for single-event entrainment mode. For the multi-event entrainment mode, the idea may be available only if merging multiple adjacent sweep events and transferring energy to the same displacement events has been further developed.

## 6. Conclusions

The intermittent creep of gravel particles, as well as particle rocking, is of importance for assessing geomorphologic work and estimating aeolian transport processes. This study develops a micro--mechanical model to characterize the incipient motion of gravel particles assumed to be initiated by coherent airflow structures (quadrant events). A work-based criterion Eq. (5), is established from the developed model to define the full entrainment of coarse particles by rolling.

The proposed criterion agrees well with observations obtained from wind tunnel experiments. On the flat and rough bed surface, it was found that both rocking and rolling are initiated primarily by sweep events capable of inducing sufficient drag force to exceed the surface resistance. Beyond that instance, rolling will occur if the aerodynamic forcing event has succificent duration and energy intensity for fully dislodging the rolling particle. Otherwise the particle's motion is classified as a rocking event, where the particle returns back to its resting pocket. These latter type of motions do not contribute to aeolian transport, other than indirectly, as is done for example in the case they are enabling mechanical sieving. The proposed criterion can be used for indirectly estimating natural coarse (such as sand or gravel) particles' rocking and rolling probabilities near threshold flow conditions, as well as directly, for the assessment of dynamic thresholds of anthropogenic pollutants of similar shape and density (such as plastic debris).

**Acknowledgments:** This research has been supported by the National Natural Science Foundation of China (Grants No. 41171005, 41071005, 12272344) and the Ministry of Science and Technology of the People's Republic of China (Grant No. 2013CB956000).

**Open Research**

**Data availability statement:** The data (Valyrakis and Xiao-Hu, 2024) used in this manuscript, as well as for the wind velocity profile and turbulence spectra are archived in a repository, and can be accessed via the following URLs: https://zenodo.org/records/7782901.